\documentclass[aps,superscriptaddress,amsmath,amssymb,floatfix,twocolumn,showpacs,amsfonts,longbibliography]{revtex4-2}
\usepackage{amsmath}
\usepackage{enumerate}
\usepackage{bm}
\usepackage[dvipdf]{graphicx}
\usepackage{floatrow}
\usepackage[caption=false]{subfig}
\usepackage[normalem]{ulem}
\usepackage{soul}
\usepackage{xr}
\makeatletter

\newcommand*{\addFileDependency}[1]{% argument=file name and extension
\typeout{(#1)}% latexmk will find this if $recorder=0
\@addtofilelist{#1}
\IfFileExists{#1}{}{\typeout{No file #1.}}
}\makeatother

\newcommand*{\myexternaldocument}[1]{%
\externaldocument{#1}%
\addFileDependency{#1.tex}%
\addFileDependency{#1.aux}%
}
\myexternaldocument{table_of_parameters}
\myexternaldocument{table_MR}
\myexternaldocument{fig0}
\myexternaldocument{fig1}
\myexternaldocument{fig2}
\myexternaldocument{fig3}

\usepackage{cleveref}
\usepackage{mathrsfs}
\usepackage{xcolor}

\begin{document}

\title{Tunable viscous layers in Corbino geometry using density junctions}

\author{Ramal Afrose}
\affiliation{Department of Physics, Faculty of Science,
National University of Singapore, Science Drive 3,
Singapore 117542}

\author{Ayd\i n Cem Keser}
\affiliation{Australian Research Council Centre of Excellence in Low-Energy Electronics Technologies,The University of New South Wales, Sydney 2052, Australia}

\author{Oleg P. Sushkov}
\affiliation{Australian Research Council Centre of Excellence in Low-Energy Electronics Technologies,The University of New South Wales, Sydney 2052, Australia}
\affiliation{School of Physics, University of New South Wales, Kensington, NSW 2052, Australia}

\author{Shaffique Adam}
\affiliation{Department of Physics, Faculty of Science,
National University of Singapore, Science Drive 3,
Singapore 117542}
\affiliation{Centre for Advanced 2D Materials,
National University of Singapore, 6 Science Drive 2, 
Singapore 117546}
\affiliation{Department of Materials Science and Engineering,
National University of Singapore, 9 Engineering Drive 1,
Singapore 117575}
\affiliation{Yale-NUS College, 16 College Avenue West,
Singapore 138527}

\date{\today}
%
%%%%%%%%%%%%%%
\begin{abstract}
In sufficiently clean materials where electron-electron interactions are strong compared to momentum-relaxing scattering processes, electron transport resembles the flow of a viscous fluid.  We study hydrodynamic electron transport across density interfaces (n-n junctions) in a 2DEG in the Corbino geometry.
From numerical simulations in COMSOL using realistic parameters, we show that we can produce tunable viscous layers at the density interface by varying the density ratio of charge carriers. We quantitatively explain this observation with simple analytic expressions together with boundary conditions at the interface.
We also show signatures of these viscous layers in the magnetoresistance. Breaking down viscous and ohmic contributions, we find that when outer radial region of the Corbino has higher charge density compared to the inner region, the viscous layers at the interface serve to suppress the magneto-resistance produced by momentum-relaxing scattering. Conversely, the magneto-resistance is enhanced when the inner region has higher density than the outer. 
Our results add to the repertoire of techniques for engineering viscous electron flows, which hold a promise for applications in future electronic devices.
\end{abstract}
%%%%%%%%%%%%%%
\maketitle

 \section {Introduction}
 \label{sec:introduction}

\noindent  In lieu of the Drude flow in conventional conductors, electrons flow like a viscous fluid when collisions among them become the dominant scattering mechanism~\cite{gurzhi_hydrodynamic_1968,gurzhi_minimum_1963,gurzhi_electron-electron_1995,Ho_hydrodynamic_window}. This hydrodynamic regime has long remained elusive in experiments due to want of fabrication of sufficiently clean materials where electron-electron interactions are strong compared to momentum-relaxing scattering. However, the advent of ultra-high mobility 2D electron systems has bridged this gap. Several effects of viscous electron flow like negative non-local resistance~\cite{Bandurin_non_local_resistance,Levin_non_local_resistance}, electron-hole drag~\cite{Tan_eh_drag_blg}, vorticity~\cite{levitov_electron_2016,aharon-steinberg_direct_2022}, Poiseulle flow~\cite{sulpizio_visualizing_2019,ku_imaging_2020}, superballistic conductance through point contacts~\cite{Guo_Levitov_superballistic,krishna_kumar_superballistic_2017,ginzburg_2023}, and violation of Widemann-Franz law~\cite{Crossno_Fong,gooth_thermal_2018,lucas_electronic_2018,jaoui_thermal_2021,ahn_mesoscale_2022} are predicted and have also been observed. With a magnetic field, more unconventional effects like negative magneto-resistance~\cite{Alekseev_negative_MR,Shi_negative_MR,Aydin,Hatke_giant_negative_MR,mani_size-dependent_2013,gusev_viscous_2021}, Hall viscosity~\cite{Gusev,Berdyugin} and giant anomalous photoresistivity~\cite{wang_hydrodynamic_2022,dai_observation_2010,hatke_giant_2011,bialek_photoresponse_2015,alekseev_transverse_2019} have been seen.

Two main geometries that have been used to study electron transport at the mesoscopic scale are the Hall bar and the Corbino ring. Unlike the Hall bar, the Corbino does not have edges except for the source and drain terminals. This distinctive feature makes it an attractive setup to study bulk states in the quantum Hall regime, since quantum Hall transport measurements in the more conventional Hall bar geometry are dominated by contribution from edge currents.  In addition, due to transverse Hall currents in a magnetic field, the Corbino makes magneto-resistance a feasible probe to study hydrodynamics~\cite{Shavit,Gall_Corbino_magnetoresistance_in_neutral_graphene,Levchenko_MR,Tomadin_Corbino_viscometer}. For example, Ref.~\cite{Shavit} derives a quadratic-in-field magneto resistance in a 2DEG Corbino ring (e.g. in GaAs heterojunctions), and shows that an applied electric field is expelled from the bulk of the sample in spite of viscous dissipation. However, at low carrier densities, due to non-vanishing temperature gradients this is no longer true~\cite{Li_2022,Gall_2023}. Ref.~\cite{Gall_Corbino_magnetoresistance_in_neutral_graphene} shows how viscosity affects magneto-resistance in charge neutral graphene Corbino ring assuming no-slip and no-stress leads.  Ref.~\cite{Levchenko_MR} extends the study to low-density and the high-density limiting cases and show that although the simple expression of Ref.~\cite{Shavit} is valid in the high-doping Fermi liquid regime, additional contributions appear near neutrality point. Thermoelectric coefficients calculated in the ballistic limit~\cite{Rycerz} show signatures of transition from quantum Hall transport to incoherent transport. Since we are only concerned with the high charge density transport regime, we use the formalism of Ref.~\cite{Shavit}.  

\begin{figure*}[th!]
    \centering
    \sidesubfloat[]{\includegraphics[width=0.15\linewidth]{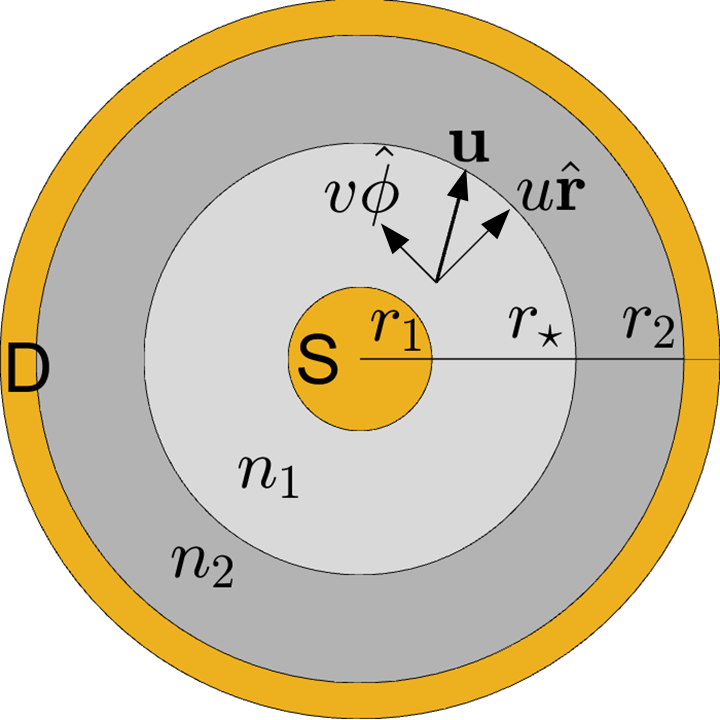}}
    \sidesubfloat[]{\includegraphics[width=0.25\linewidth]{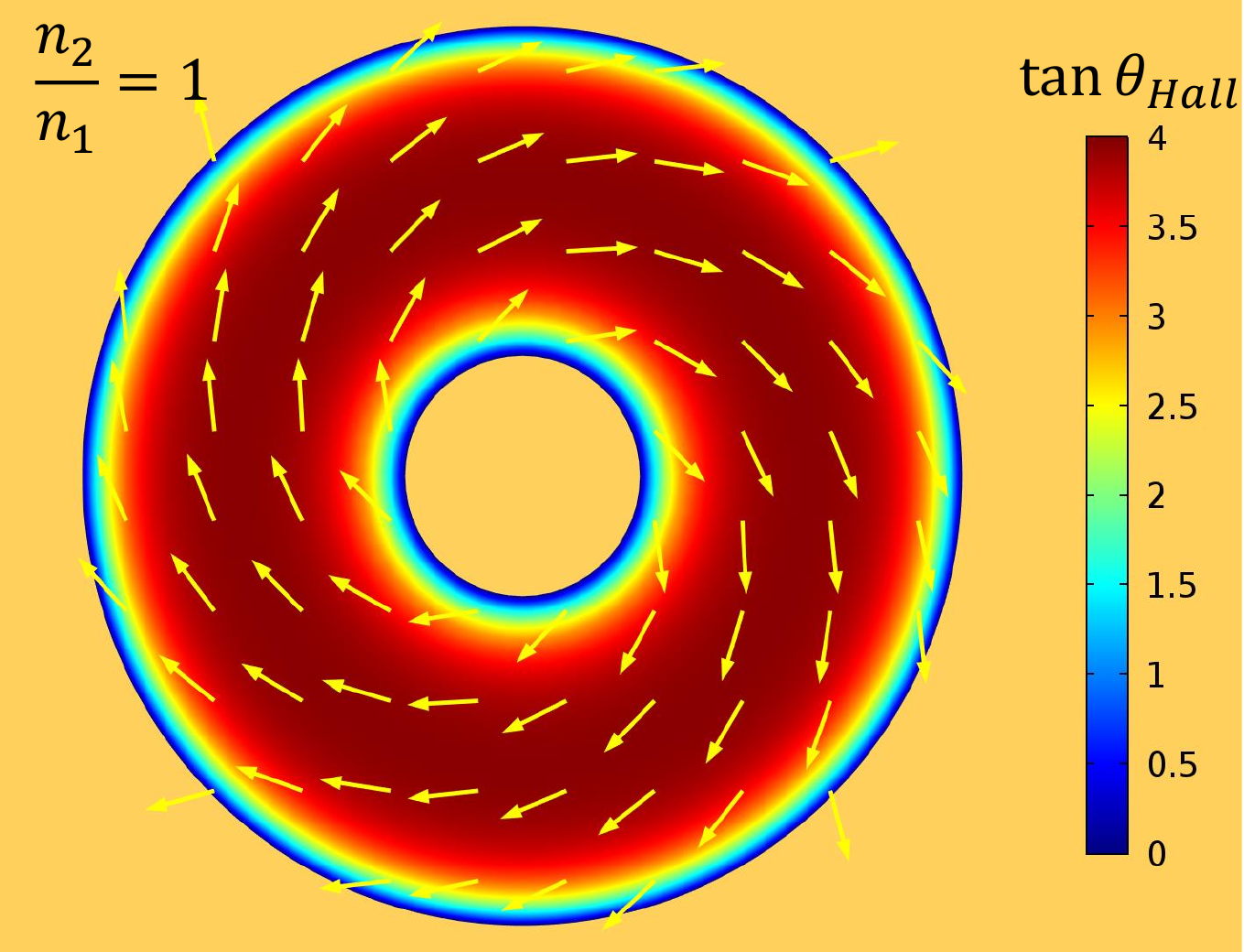}}
    \sidesubfloat[]{\includegraphics[width=0.25\linewidth]{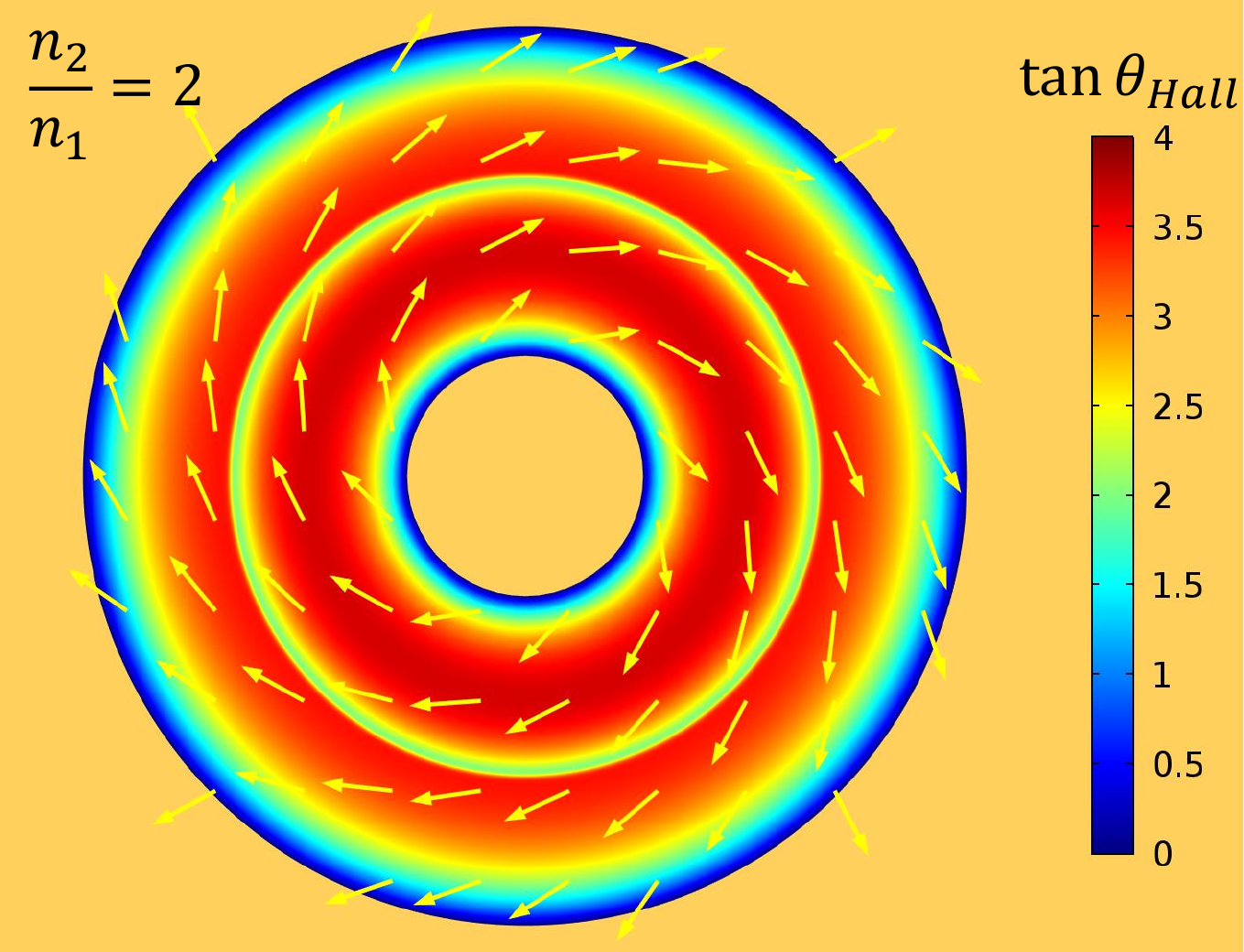}}
    \sidesubfloat[]{\includegraphics[width=0.25\linewidth]{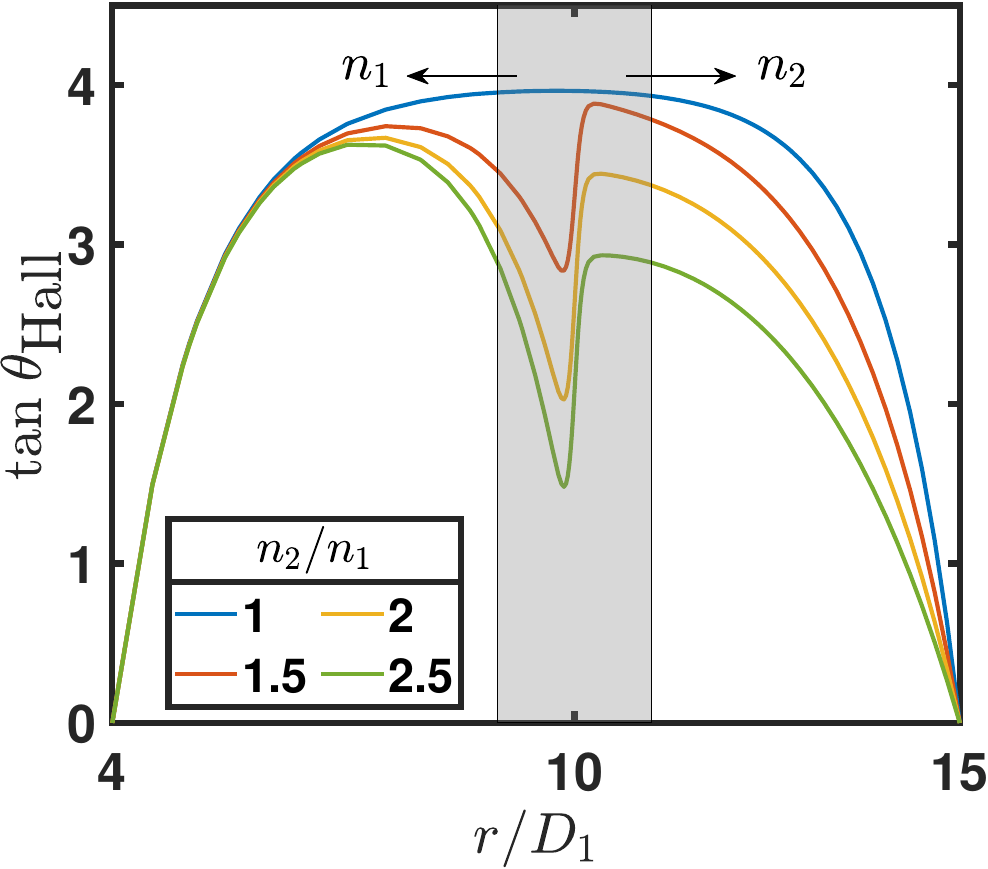}}
    \caption{(a) Schematic of the setup considered in this paper. (b), (c) Colormap of $\tan \theta_{\text{Hall}}=v(r)/u(r)$ for uniform density ($n_2/n_1=1$), and a density gradient $n_2/n_1=2$, obtained from a numerical simulation of the hydrodynamic equations (\ref{hydro_eqns}) in COMSOL. The Hall angle is expected to be approximately constant in the bulk where flow is Drude-like, and change towards the contacts due to drag forces acting on the electrons due to viscous layers. A density gradient induces formation of additional viscous layers at the interface, sensitive to the density gradient. (d) Radial profile (along a radial section of (c)) of $\tan \theta_{\text{Hall}}$ for various outer densities $n_2$. The interface induces a change in Hall angle in a region (shaded) approximately wide as the Gurzhi length $D$ on either side. The jump in Hall angle at the interface also increases for a higher density gradient.}
    \label{fig:0}
\end{figure*}

The situation is very different in the Hall bar geometry, where magneto resistance is either very weak (for small fields) or is complicated by several factors such as change in viscosity~\cite{Alekseev_negative_MR,Gusev}, size of cyclotron orbit compared to channel width~\cite{Moore,Holder_MR_Hall_bar}, edge currents~\cite{panchal_visualisation_2014}, etc. Moreover, the presence of edges introduces an unknown boundary condition which effects the flow~\cite{Kiselev_boundary_conditions}. Recent efforts have been made to mitigate this problem by making samples with perfect-slip boundaries. This was done by inducing an electron channel in a GaAs heterostructure by applying a bias from a top gate instead of chemical doping. There, viscous effects were artificially re-introduced by modifying the geometry of the channel~\cite{Aydin}, or by means of magnetic modulation~\cite{Engdahl_micromagnets,keser_magnetic_2023,engdahl_2023}. The Corbino geometry provides an alternative solution to the problem of boundary conditions by completely eliminating boundaries except for contacts at source and drain.

Experiments on the Corbino in graphene have directly probed the existence of viscous layers localised at the input and output terminals~\cite{kumar_imaging_2022}. These layers are approximately as wide as the `Gurzhi length' $D$,  that is, the momentum diffusion length arising from combination of momentum relaxing and momentum conserving e-e scattering, and defined as $D = \sqrt{l_{\rm mr} l_{\rm ee}}/2$. Due to the high mobility of 2DESs with respect to the metallic leads, and due to rough surfaces of the leads over microscopic length scales, electron flow into and out of the sample is almost always normal. In this Article, we explore the possibility of a situation where the injected velocity has a tunable tangential component, by considering electron flow through two regions of different density (n-n junction) in the Corbino geometry. Away from the boundaries, viscous forces cease to operate and the electron transport assumes  Drude character. The Drude flow depends on the charge density of carriers and is different on either side of the junction. At a no-stress interface of two densities, the tangential flow must be continuous, forcing the velocity to reduce to a common value and leading to formation of viscous layers. The strength of these layers is directly proportional to the mismatch between the interface velocity and the Drude velocity in the bulk. Therefore, by tuning the gate bias, we can easily realise viscous layers of varying lengths.

\par Manipulating viscous electron flow at mesoscopic length scales is an actively pursued endeavour. Recent works studying hydrodynamic flows through different channel geometries~\cite{Levin_geometric_engineering} find that effective channel width can change for different geometries while the microscopic scattering parameters are unaffected.
Although boundary conditions are never perfectly determined microscopically, a perfect no-slip boundary can be realised over larger length scales by considering current flowing through a series of constrictions~\cite{Moessner_geometric_engineering}. Ref.~\cite{Kiselev_boundary_conditions} show that slip length at the boundaries can vary with temperature in a non-trivial way, while other studies~\cite{Hui_nonlinear_hydrodynamics} demonstrate how non-linear hydrodynamic effects like Bernoulli effect, Eckart streaming and Rayleigh streaming can be realised in special scenarios.  Our proposal of a gate-tunable viscous layer could be used for easy electrical manipulation of thermal dissipation at interfaces and adds to the growing repertoire of methods for viscous flow engineering.

\par The plan of the paper is as follows. In Sec. \ref{sec: tunable viscous layers}, we show the emergence of distinct viscous layers at a density interface from numerical simulation of compressible Stokes flow in COMSOL.  We next present an analytical explanation of this result in Sec. \ref{sec:analytic_derivation} using simple expressions derived from the more complicated exact solution. We show that we can piecewise model the flow and match them at the interface by using simple interface conditions, which we derive in Sec \ref{sec:Bdy cond}. Then, in Section \ref{sec: magnetoresistance}, we show signatures of these viscous layers in the magneto resistance. We also decompose the ohmic and viscous contributions with analytic expressions. Finally, we end with a summary of our findings in Section \ref{sec:summary}.

\begin{table*} \label{table:parameters}
\caption{Value of parameters of 2DEG used for simulation of hydrodynamic flow. {Parabolic dispersion 
$\epsilon = p^2/(2 m^*)$, with spin degeneracy $g_s = 2$ is assumed.}}
\begin{centering}
\begin{tabular}{|c |c |c||c |c |c|}
\hline 
Parameter & Description & Value & Parameter & Description & Value\tabularnewline
\hline 
\hline 
$n_1$ & charge carrier density & $2\times10^{11}\text{cm}^{-2}$ & $m^*$ & effective electron mass & $0.067m_e=6.1\times 10^{-32}\text{kg}$\tabularnewline
% \hline 
$B$ & Applied magnetic field & $20\text{mT}$ & $l_{\text{ee}}$ & e-e scattering length for $n=n_1$ & $300\text{nm}$\tabularnewline
% \hline 
$\mu$ & mobility & $2\times10^{6}\text{cm}^{2}/\text{V}\cdot\text{s}$ & $l_{\text{mr}}$ & momentum-relaxing scattering length & $14.8\mu\text{m}$\tabularnewline
% \hline 
$D_1$ & Gurzhi length for $n=n_1$ & $1.05\mu\text{m}$ & 2$d$ & width of interface & $0.2D_1=210\text{nm}$\tabularnewline
$r_1$ & radius of inner contact & $4D_1=4.2\mu\text m$ & $r_2$ & radius of outer contact & $15D_1=15.8\mu\text m$\tabularnewline
\hline 
\end{tabular}
\par\end{centering}
\end{table*}

\section{Tunable viscous layers at density interface} \label{sec: tunable viscous layers}

In this section, we present results from numerical simulation of hydrodynamic flow across a density gradient and show that viscous layers can be induced at the interface of different density regions. In the parameter window favouring hydrodynamic regime of transport, the local conservation laws of particle number and momentum which govern the electron flow take the form of the equation of continuity and the Stokes equation~\cite{Aydin} 
\begin{subequations}
\label{hydro_eqns}
\begin{align}
\nabla\cdot\left(n\mathbf{u}\right)  &=0 \label{eq:continuity}\\
ne\left(\mathbf{E}+\mathbf{u}\times\mathbf{B}\right)+&\nabla\cdot\boldsymbol{\sigma}-\frac{m^{*}n}{\tau}\mathbf{u}  =0\label{eq:Stokes}
\end{align}
\end{subequations}
 where $\mathbf u$ is the macroscopic velocity of fluid elements, $\boldsymbol{\sigma}=\eta\left[\nabla\otimes\mathbf{u}+\left(\nabla\otimes\mathbf{u}\right)^{T}-\left(\nabla\cdot\mathbf{u}\right)I\right]$ is the shear stress tensor. We have neglected pressure gradients because, assuming the `gradual channel approximation', they result in a negligible correction to the capacitance between the 2DEG and the gates, and can be absorbed into the electric potential. The bulk viscosity of electrons is considered negligible~\cite{Principi_bulk_viscosity,Aydin}. Equation (\ref{hydro_eqns}) is supplemented by boundary conditions, which are determined by the magnitude and direction of current flowing in and out of the 2DEG at the inner and outer electrodes.
 
 We also omitted the nonlinear convective derivative (Navier term) $\mathbf{u}\cdot\nabla \mathbf{u}$ on account of the very low Reynolds number ($<0.1$) in the electron fluid. At temperatures when hydrodynamic effects become prominent, thermal motion of electrons become sufficiently strong that effects of Landau quantization are negligible.  Furthermore, the Hall viscosity compares to the shear viscosity by a factor of $2\omega_c \tau_{ee}$~\cite{Alekseev_negative_MR}.  From values in Table \ref{table:parameters}, this is $\approx 0.16$, which we assume is small and therefore, negligible.
 
 For the case of homogeneous charge density and assuming azimuthal symmetry, with the convention $\mathbf{u}=u\hat{\mathbf{r}}+v\hat{\boldsymbol{\phi}}$, the exact solution
of (\ref{hydro_eqns}) is given by~\cite{Gall_2023,levchenko_transport_2020}
\begin{subequations}
\label{eq:soln_Stokes}
    \begin{gather}
        u(r) = u_1r_1/r\\
        v\left(r\right) =c_{1}I_{1}\left(r/D\right)+c_{2}K_{1}\left(r/D\right)-\omega_{c}\tau u_{1}r_{1}/r
\end{gather}
\end{subequations}
 where $\omega_{c}=eB/m^{*}$ is the cyclotron frequency, $D=\sqrt{\nu\tau}$
is the Gurzhi length, $I_{1},K_{1}$ are first-order modified Bessel
functions, $c_{1},c_{2}$ are constants determined by fitting to boundary
conditions, and $u_{1}=u\left(r_{1}\right)$ is the input
radial velocity.

Comparing this with the non-viscous, simple Drude solution
\begin{equation}
v^{\mathsf{ohm}}\left(r\right)=-\omega_{c}\tau u_{1}\frac{r_{1}}{r}
\label{eq:Drude_soln}
\end{equation}
we can see that viscosity conspires with momentum-relaxing scattering in the form of the Gurzhi length to affect the tangential velocity $v$. A simple way to highlight this effect is to calculate the local angle between the flow velocity and the radial vector, known as the Hall angle: $\tan \theta_{\text{Hall}}=v(r)/u(r)$. For Drude flow (Eq. (\ref{eq:Drude_soln})), this is constant, while viscous terms in Eq. (\ref{eq:soln_Stokes}) cause deviations from it. At the source and drain, where flow is approximately normal to the surface of the leads, the Hall angle is zero, therefore, viscous layers develop near these terminals to accelerate the tangential flow.

In our case, we consider a non-uniform density profile $n(r)$ varying over a relatively small length scale $d\ll D$ at an interface at $r=r_{\star}$. In a gated junction in GaAs quantum wells, the density interface is expected to have a width $d\approx 100~\text{nm} -150~\text{nm}$. On solving the electrostatic potential due to the gates, we find the density varies as
\begin{equation}
\label{eq:density}
    n(r) = \frac{n_1+n_2}{2} + \frac{n_2-n_1}{2} \tanh([r-r_\star]/d)
\end{equation} 
which interpolates between the two regions $r <r_\star $ and $r>r_\star$ with densities $n_{1},n_2$ respectively.

Given this density profile, we solve the hydrodynamic equations (\ref{hydro_eqns}) in the numerical solver COMSOL, supplemented by no-slip conditions at the source and drain. We assume a  density gradient in Eq. (\ref{eq:density}) that smoothly connects the two constant density regions over a characteristic length of {$d=0.1D \approx 100\:\text{nm}$}.
Such a density setup can be created by using a dual gate architecture with top and bottom gates~\cite{Elahi}. 
% The thickness of the interface is almost the same as the distance of the gate from the 2DEG.
We assume Fermi-liquid behaviour of electrons with respect to density, i.e, the e-e scattering rate goes as {$\tau_{ee}^{-1} \propto 1/E_F\propto 1/n$}. Also, given the high mobility of 2DEG, we assume the momentum-relaxing scattering is limited by phonons and not by disorder, therefore, $\tau$ is independent of $n$~\cite{Aydin, arora_phonon-scattering-limited_1985, ridley_electron-phonon_1982, kawamura_temperature_1990}. 
The Gurzhi length then varies as $D\propto n$. At hydrodynamic temperatures $l_{ee}\sim n$~\cite{Aydin} or even stronger, therefore qualitatively $D\sim n$ holds for a large range of parameters. We have assumed an electron-electron scattering length $l_{ee}=300\text{nm}$ at $n_1=2\times10^{11}\text{cm}^{-2}$, which corresponds to a temperature $T=20\text{K}$ in 2DEG in GaAs/AlGaAs~\cite{Aydin}. {The Fermi temperature is $T_F=83\: \text K$, and the Fermi wavenumber and velocity, $k_F=(9 \text {nm})^{-1}$, $v_F=1.9\times 10^7 \text{cm/s}$, assuming a parabolic dispersion of the conduction band.} A summary of the values of parameters is given in Table \ref{table:parameters}. 

In Fig. (\ref{fig:0}), we plot the Hall angle profile for uniform density and a finite density gradient $n_2=2n_1$. We clearly see a change in the Hall angle near the density interface, indicating the formation of viscous layers. We also show the radial profile of $\tan \theta_{\text{Hall}}$ for different density gradients. We find that a larger gradient causes a greater change in Hall angle, thereby creating stronger viscous layers.

{The appearance of viscous boundary layers has already been shown in a graphene Corbino ring by mapping the Hall angle profile using single-electron transistor imaging}~\cite{kumar_imaging_2022}. Other techniques like nitrogen-vacancy center magnetometry have been used to map flow profiles of electrons in mesoscopic systems~\cite{Jenkins2022}. In light of such developments, we believe interface-induced viscous layers can also be observed by imaging the flow profile in experiments.
\section{Analytic derivation of tunable viscous layers}\label{sec:analytic_derivation}

The hydrodynamic equations (\ref{hydro_eqns}) for homogeneous charge density are solved by Eqs. (\ref{eq:soln_Stokes}). This form, although known in literature, is very non-intuitive. We show in appendix \ref{app:Approximate_solution} that it can be approximated by the much simpler expression
\begin{align}
v\left(r\right)=&\left(\omega_{c}\tau u_{1}+v_{1}\right)\sqrt{\frac{r_{1}}{r}}e^{-\left(r-r_{1}\right)/D}-\omega_{c}\tau u_{1}\frac{r_{1}}{r} \nonumber \\
&+\left(\omega_{c}\tau u_{2}+v_{2}\right)\sqrt{\frac{r_{2}}{r}}e^{-\left(r_{2}-r\right)/D},\qquad r_{2}>r_{1}\gtrsim D\label{eq:approx soln}
\end{align}
where $r_1, r_2$ are the radii of the inner and outer contacts. This shows that there are two viscous layers exponentially localized
over a length $D$ at the inner and outer terminals at $r_{1},r_{2}$, and a Drude contribution which dominates in the bulk $r_{1}+D\lesssim r\lesssim r_{2}-D$. Surprisingly, if the injected velocity $v_{1}$ matches the non-viscous Drude value $v^{\mathsf{ohm}}(r_1)$ in (\ref{eq:Drude_soln}), Eq.~(\ref{eq:approx soln})
predicts that the viscous layers disappear completely. As
noted earlier, metal-2DEG interfaces are almost always no-slip ($v_{1},v_{2}\approx0$),
so this situation is never realized.

For flow through a density gradient, the continuity equation constrains the radial velocity as $u(r)=n_1r_1u_1/n(r)r$. {The bulk ohmic velocity $v^{\textsf {ohm}}\sim -\omega_c\tau u(r)$ is therefore discontinuous at the interface, while the boundary condition} (\ref{eq:BC v}) {derived in the next section states that the net tangential velocity $v$ must be continuous.} Therefore, viscous layers must develop at the interface at $r_{\star}$ to force $v$ to a common value. Moreover, the viscous dissipation in these layers is proportional to the mismatch between the Drude velocity in the bulk and the interface velocity. By tuning the density ratio, we can tune this mismatch and thereby produce viscous layers of varying strengths.

\begin{figure}
    \centering
    \sidesubfloat[]{\includegraphics[width=0.6\columnwidth]{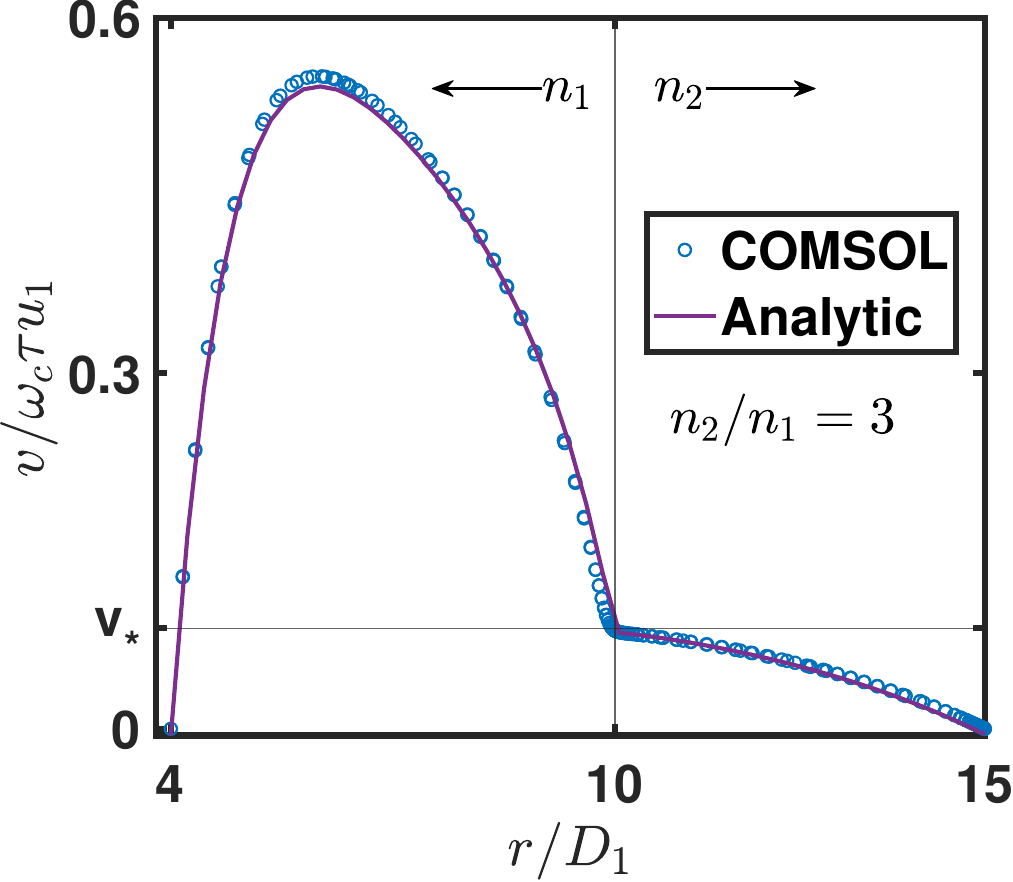}} \label{fig:vfit}
    \sidesubfloat[]{\includegraphics[width=0.6\columnwidth]{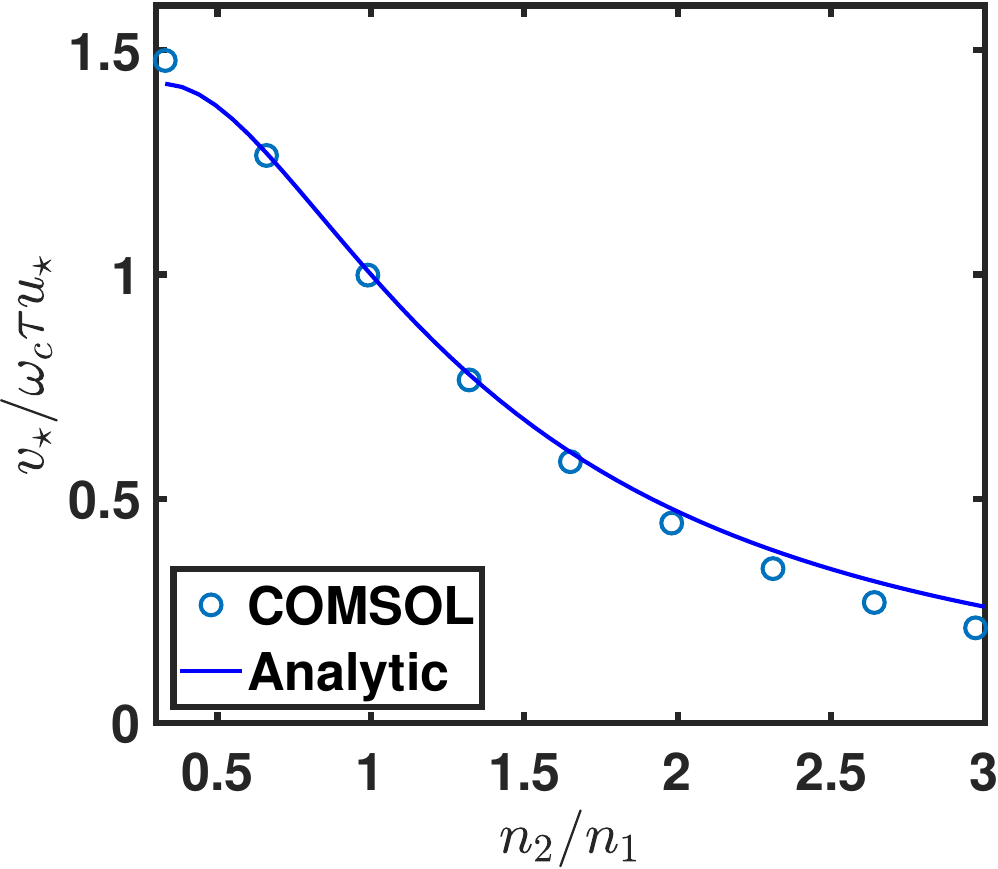}}
    \caption{(a) Plot of tangential flow velocity as function of radial distance from eq. (\ref{eq:vfit}). Numerical simulation of Eq.~\eqref{hydro_eqns} using COMSOL with smooth density gradient as in Eq.~\eqref{eq:density} with $d = 0.1D_1~\sim 100\:\text{nm}$ shows that the piecewise analytical solution with interface conditions approximates the flow very well. Viscous layers approximately wide as the Gurzhi length $D$ act near the input, output terminals and at the interface to match the bulk Ohmic velocity to flow at the boundaries. Tuning the density ratio changes the tangential velocity at the interface, $v_{\star}$, which, in turn modulates the strength of viscous layers. (b) Velocity at interface as function of density ratio, obtained using analytic solution (\ref{eq:vfit}) and using boundary conditions (\ref{eq:BC sig_rphi}), (\ref{eq:BC v}). $v_{\star}$ can be made to vary by a factor of 3 by varying $n_2/n_1$ over experimentally feasible ranges. COMSOL simulation of the Stokes flow with smooth interface gives identical values of $v_{\star}$ and verifies that our interface conditions are accurate. 
    % Inset: Schematic of the Corbino setup. Notation used for velocity components are shown.
    }
    \label{fig:1}
\end{figure}

In Fig (\ref{fig:1}), we plot the velocity profile as a function of the radial
distance $r$ using the fit solution
\begin{align}
v\left(r\right) &=\left(\omega_{c}\tau u_{1}+v_{1}\right)\sqrt{\frac{r_{1}}{r}}e^{-\left(r-r_{1}\right)/D_{1}}-\omega_{c}\tau u\left(r\right) \nonumber \\
&+v^{\mathsf{int}}\left(r\right)+\left(\omega_{c}\tau u_{2}+v_{2}\right)\sqrt{\frac{r_{2}}{r}}e^{-\left(r_{2}-r\right)/D_{2}}\label{eq:vfit}
\end{align}
 where the velocity near the interface ($u_{\star}\equiv u(r_{\star}-d) = u_1r_1/r_{\star}$)
\begin{align*}
\renewcommand{\arraystretch}{1.5}
    v^{\mathsf{int}}\left(r\right)=
    \begin{cases}
        \left(\omega_{c}\tau u_{\star}+v_{\star}\right)\sqrt{\frac{r_{\star}}{r}}e^{-\left(r_{\star}-r\right)/D_{1}} & r<r_{\star}\\
        \left(\omega_{c}\tau u_{\star}\frac{n_{1}}{n_{2}}+v_{\star}\right)\sqrt{\frac{r_{\star}}{r}}e^{-\left(r-r_{\star}\right)/D_{2}} & r>r_{\star}
    \end{cases}
\end{align*}

\par We support this result with a numerical simulation of the Stokes
flow in COMSOL. As it can be seen, our approximate solution bears excellent agreement
with the exact result. Moreover, from boundary condition (\ref{eq:BC sig_rphi}), we
can equate the off-diagonal stress tensor $\sigma_{r\phi}$ at $r_{\star}$
to get the common velocity $v_{\star}$.
\begin{equation}
-\frac{v_{\star}}{\omega_{c}\tau u_{\star}}=\frac{1+2\rho_{\star}+2\rho_{\star}\overline{n}-\overline{n}^{2}}{-3+2\rho_{\star}+2\rho_{\star}\overline{n}^{2}+3\overline{n}^{3}}
\label{vstar}
\end{equation}
where $\rho_{\star}=r_{\star}/D_{1}$ and $\overline{n}=n_{2}/n_{1}$.
A plot of $v_{\star}$ vs $\overline{n}$ is also shown. Numerical simulation of the flow across a smooth density gradient of width $d\ll D$ produces very similar results to the piece wise analytical solution matched with interface conditions.

Thus, we see that by changing the density ratio, we can tune the velocity
$v_{\star}$ at the interface, by which we can control the
appearance of viscous layers. This also has implications in the electric
resistance, as we explore in Section \ref{sec: magnetoresistance}.

\section{Boundary conditions of flow at interface}\label{sec:Bdy cond}

We derive the boundary conditions on which our previous results are based. We consider a sharp interface between two regions of densities $n_{1}$ and $n_{2}$. By sharp, it is implied that the variation of density
is over length $d$ much smaller than the dimensions of the sample,
but larger than the Fermi wavelength. To derive boundary conditions, we integrate the
equations of flow (\ref{eq:continuity}), (\ref{eq:Stokes}) over a patch with faces parallel to the interface. Because of the $\phi$ symmetry of the system, this is the same as integrating
the equations from $r_{\star}-d$ to $r_{\star}+d$, where $2d$ is the thickness of the interface. 

For the continuity equation,
\[
\int_{r_{\star}-d}^{r_{\star}+d}dr\;\partial_{r}\left(rnu\right)=0
\]
The radial velocity is therefore discontinuous, as 
\begin{equation}
\left(nu\right)\big|_{r_{\star}-d}^{r_{\star}+d}= 0\label{eq:u(r)}
\end{equation}

Integrating the tangential component of (\ref{eq:Stokes}),
\begin{align*}
-eB\int_{r_{\star}-d}^{r_{\star}+d}rdr\;nu+\int_{r_{\star}-d}^{r_{\star}+d}dr\frac{\partial_{r}\left(r^{2}\sigma_{r\phi}\right)}{r} & \\
-\frac{m^{\star}}{\tau}\int_{r_{\star}-d}^{r_{\star}+d}rdr\;nv &=0
\end{align*}
 From the continuity equation, $r\;nu=\mathrm{const}$, so, in the
limit of small $d$, the first term gives a vanishing contribution.
Similarly, assuming the tangential flow $v$ does not diverge at the
interface, the third term contributes negligibly in limit of
small $d$. Therefore,
\begin{align*}
0\approx\frac{1}{r_{\star}}\int_{r_{\star}-d}^{r_{\star}+d}dr\;\partial_{r}\left(r^{2}\sigma_{r\phi}\right)=\frac{1}{r_{\star}}\left(r^{2}\sigma_{r\phi}\right)\bigg|_{r_{\star}-d}^{r_{\star}+d}
\end{align*}
 In other words, the off-diagonal stress tensor, $\sigma_{r\phi}$,
is continuous at the interface:
\begin{equation}
\sigma_{r\phi}\big|_{r_{\star}-d}^{r_{\star}+d}=0\label{eq:BC sig_rphi}
\end{equation}

Given $\sigma_{r\phi}=r\partial_{r}\left(v/r\right)$, this
implies that the tangential flow velocity $v$ is also continuous.
\begin{equation}
v\big|_{r_{\star}-d}^{r_{\star}+d}=0\label{eq:BC v}
\end{equation}

Multiplying the radial component of (\ref{eq:Stokes}) by $u$ and integrating,
\begin{equation}
\begin{split}
-e\int_{r_{\star}-d}^{r_{\star}+d}rdr\;nu\partial_{r}\Phi+eB\int_{r_{\star}-d}^{r_{\star}+d}rdr\;nuv \\
+\int_{r_{\star}-d}^{r_{\star}+d}dr\frac{u}{r}\partial_{r}\left(r^{2}\sigma_{rr}\right)-\frac{m^{*}}{\tau}\int_{r_{\star}-d}^{r_{\star}+d}rdr\;nu^{2} & =0
\end{split}
\end{equation}
 {where $\Phi$ is the electric potential.} Using the fact that $rnu=I/2\pi e=\text{const}$ across the interface, as before, we find that the terms proportional to $B,\tau^{-1}$ have a vanishing contribution in the limit of small $d$. The net condition reduces to
\begin{gather}\label{eq:radial_cond(1)}
\Phi\big|_{r_{\star}+d}^{r_{\star}-d}\times\frac{I}{2\pi}=\left(ru\sigma_{rr}\right)\bigg|_{r_{\star}+d}^{r_{\star}-d}+\int_{r_{\star}-d}^{r_{\star}+d}r\;dr\;\frac{\sigma_{rr}^{2}}{\eta}
\end{gather}

where integration by parts has been used for the right hand side. Given $\sigma_{rr}=\eta r\partial_{r}\left(u/r\right)$, we find
\begin{align}  \left(ru\sigma_{rr}\right)\bigg|_{r_{\star}+d}^{r_{\star}-d}=\left(\frac{I}{2\pi e}\right)^{2}\frac{2}{r_{\star}^{2}}\left(\frac{\eta_{2}}{n_{2}^{2}}-\frac{\eta_{1}}{n_{1}^{2}}\right)
\end{align}

Together with potential jump at the inner and outer contacts, this covers viscous dissipation in the homogeneous regions. The remaining term in Eq. (\ref{eq:radial_cond(1)}) is just the viscous dissipation due to compressive flow at the interface. Neglecting derivative of $r$ compared to $n$ at the interface,
\begin{align*}
\int_{r_{\star}-d}^{r_{\star}+d}r\;dr\;\frac{\sigma_{rr}^{2}}{\eta}	&=\left(\frac{I}{2\pi e}\right)^{2}\int dr\;{r^{3}}\eta\left(\frac{d}{dr}\frac{1}{nr^{2}}\right)^{2} \\
	&\approx\left(\frac{I}{2\pi e}\right)^{2}\int\frac{\eta}{n^{4}r}\left(\frac{dn}{dr}\right)^{2}dr
\end{align*}
Finally, in the limit $d\to 0$, we can approximate
\begin{align}
\left(\frac{dn}{dr}\right)^{2}=\left(\frac{\Delta n}{2d}\right)^{2}\text{sech}^{4}\left(\frac{r-r_{\star}}{d}\right)\approx\frac{\left(\Delta n\right)^{2}}{3d}\;\delta\left(r-r_{\star}\right)
\end{align}

Using this in the integral for viscous dissipation, the total potential drop at the interface is
\begin{align}\label{eq:BC Phi}
-\frac{\Delta\Phi^{\mathsf{int}}}{I}=&\frac{\eta_{2}}{\pi\left(n_{2}e\right)^{2}r_{\star}^{2}}-\frac{\eta_{1}}{\pi\left(n_{1}e\right)^{2}r_{\star}^{2}} 
\nonumber\\
&+\frac{\eta_{\star}}{\pi\left(n_{\star}e\right)^{2}}\frac{1}{6r_{\star}d}\left(\frac{\Delta n}{n_{\star}}\right)^{2},
\end{align}
where $\Delta n = n_2-n_1$ is the difference in densities,  $n_{\star}=(n_1+n_2)/2$ is the density in the middle of the junction, and {$\eta_{\star}=m^*n_{\star}\times v_F(n_\star)l_{ee}(n_{\star})/4$} is the corresponding shear viscosity. 
\par Eqs. (\ref{eq:u(r)}), (\ref{eq:BC sig_rphi}), (\ref{eq:BC v}) and
(\ref{eq:BC Phi}) are the required boundary conditions of the problem.

\begin{figure*}[!ht]
    \centering
    \sidesubfloat[]{\includegraphics[width=0.3\linewidth]{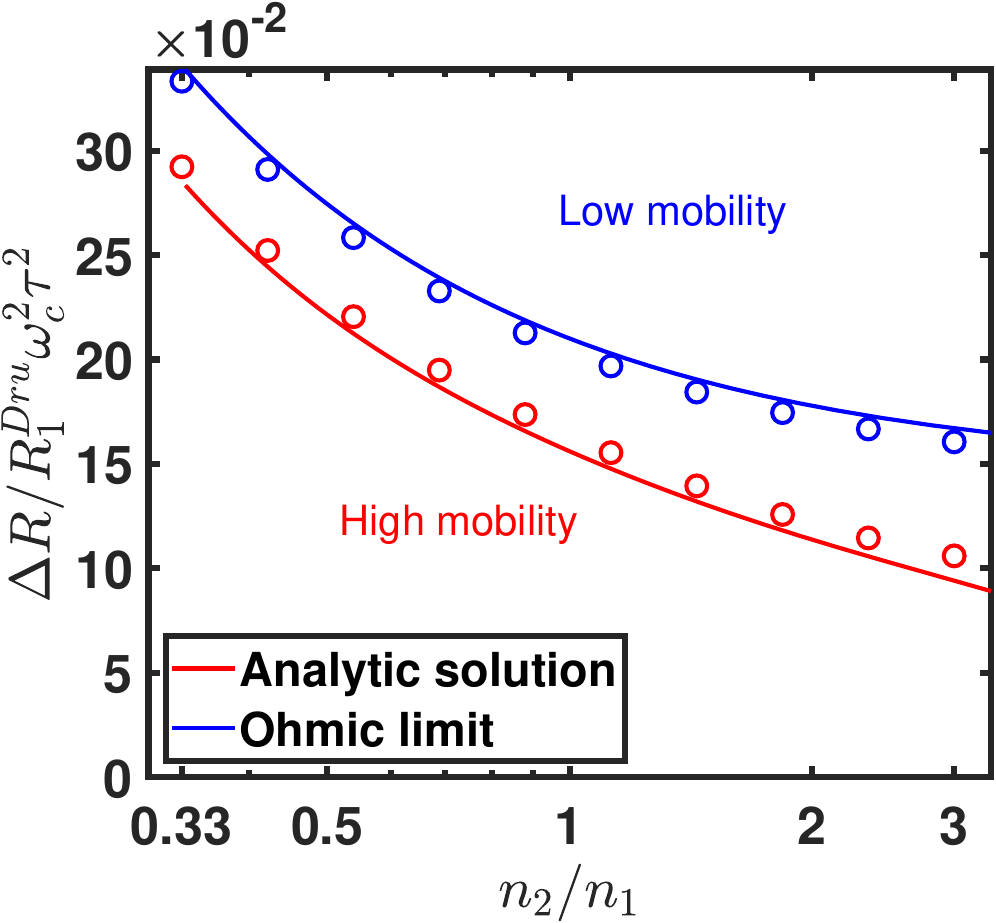}}
    \sidesubfloat[]{\includegraphics[width=0.3\linewidth]{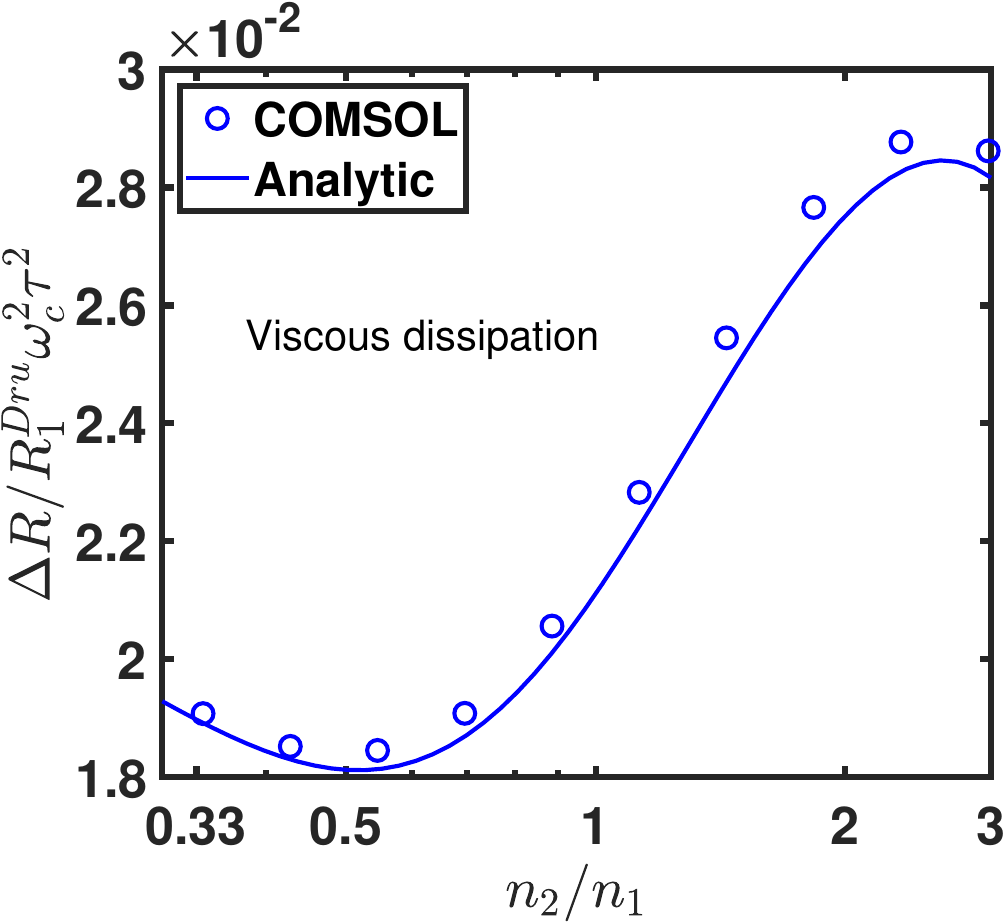}}
    \sidesubfloat[]{\includegraphics[width=0.3\linewidth]{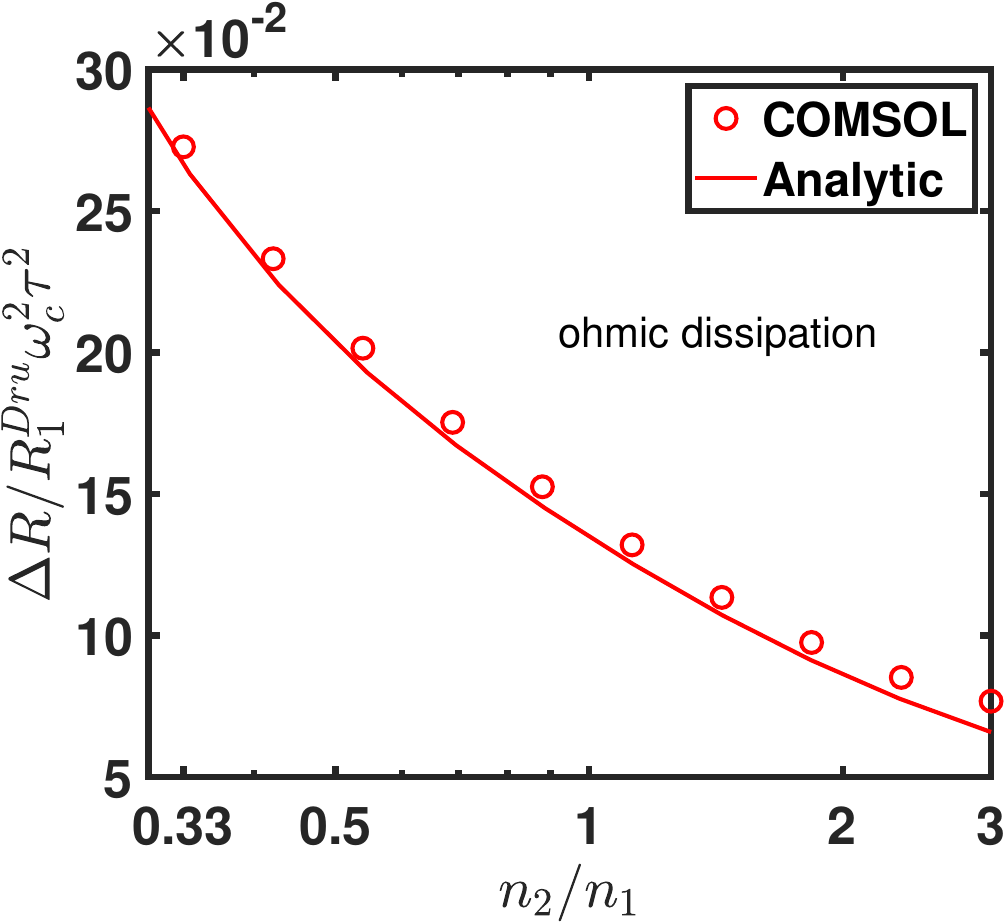}}
    \caption{(a) Scaled magnetoresistance $\left[R(B) - R(0)\right]/(R_1^\textsf{Dru}\omega_c^2\tau^2)$ with density ratio. The circles are numerical simulation values from COMSOL, obtained by calculating the potential difference between inner and outer electrodes. The red solid line is an analytical plot of sum of viscous and ohmic dissipation (see eq. (\ref{eq:power diss})). For non-viscous flow, scaled magneto resistance for different mobilities collapse onto a single curve (the Ohmic limit, eq. (\ref{eq:ohmic lim})). For high-mobility sample, viscous layers at the terminals and at interface suppress magnetoresistance (red curve). $x$ axis is set to log scale to give equal emphasis to the cases $n_2>n_1$ and $n_2<n_1$. (b) Resistance due to viscous dissipation, calculated from COMSOL and from analytic solution (\ref{eq:vfit}) (expression in table \ref{table}). Because viscosity is proportional to density, viscous dissipation goes up with $n_2/n_1$. At very high $n_2$, velocity gradient in the bulk is low, resulting in a decrease in viscous resistance. The opposite happens for very low $n_2$. (c) Resistance due to Ohmic dissipation. In the bulk, the magnitude of current density, $|J|\sim J^{\mathsf{in}}(1+\omega_c^2\tau^2)$ is approximately constant, so the ohmic resistance per unit area $R^{\mathsf{ohm}}=J^2/\mu n e$ goes down monotonically with $n_2$.}
    \label{fig:2}
\end{figure*}

\section{Signature of viscous layers in magnetoresistance}
\label{sec: magnetoresistance}

In Fig (\ref{fig:2}), we plot the magneto-resistance $\Delta R=R\left(B\right)-R\left(0\right)$
vs. $n_{2}/n_{1}$, calculated numerically for a high mobility $\mu=2\times10^{6}\text{cm}^{2}/\text V\cdot \text s$
and a low mobility $\mu=2000\;\text{cm}^{2}/\text V\cdot \text s$ Corbino ring. Assuming a ultra-high mobility sample where momentum-relaxing scattering is due to phonons only, the mobility should be close to our assumed high mobility value at $T\approx 20\:\text K$~\cite{Aydin}.
It must be noted that in the Corbino, {voltage applied divided by current
yields the inverse of magneto-conductance, which
is not equal to the magneto-resistance as the conductivity
tensor in a magnetic field is non-diagonal.} However, in the following, we use inverse
magneto-conductance and magneto-resistance interchangeably for convenience.

For non-viscous flow, described by the Drude equation, the resistance
for the density junction is simply
\begin{equation}
\Delta R^{\mathsf{ohm}}=\left(\omega_{c}\tau\right)^{2}\frac{R_{1}^{\mathsf{Dru}}}{2\pi}\left[\ln\left(\frac{r_{\star}}{r_{1}}\right)+\frac{n_{1}}{n_{2}}\ln\left(\frac{r_{2}}{r_{\star}}\right)\right] \label{eq:ohmic lim}
\end{equation}
where the Drude resistivity $R_{1}^{\mathsf{Dru}}=m^{*}/n_{1}e^{2}\tau=15.3\Omega$ for values of parameters in Table \ref{table:parameters}.
When scaled by $\left(\omega_{c}\tau\right)^{2}R_{1}^{\mathsf{Dru}}(=244\Omega)$, the
magneto resistance vs $n_{2}/n_{1}$ for samples with different mobility
should collapse onto a single curve. This is clearly reflected in
the result for the low mobility simulation in Fig (\ref{fig:2}). However, we
find that the presence of viscous layers serve to suppress this magneto-resistance.
This is counter intuitive, given that viscous dissipation increases,
but by decelerating the velocity at the interface, the viscous layers reduce
the ohmic dissipation as well.

To describe this analytically, we start with the power dissipation
for hydrodynamic flow:
\begin{equation}
I\Delta\Phi^{\textsf{bulk}}+\oint\mathbf{u}\cdot\boldsymbol{\sigma}\cdot d\mathbf{S}=\frac{1}{\tau}\int m^{*}n\mathbf{u}^{2}dV+\int\frac{\boldsymbol{\sigma}^{2}}{2\eta}dV
\label{eq:power diss}
\end{equation}
The first term on the left is the rate of work done by the electric potential
to drive the current, while the second term is the rate of work done against the boundary stress. The energy provided by these terms are dissipated by Ohmic and viscous
forces, described by the terms on the right hand side. The importance
of boundary stress is apparent if we consider the particular case of pure radial flow with no ohmic dissipation~\cite{Shavit}.
It can be shown that in this case, the potential drop $\Delta\Phi^{\textsf{bulk}}=0$,
whereas power dissipation by viscous forces is finite. The only way
the dissipated power can be compensated is by the boundary term on
the left hand side. Here, we expand this to a broader framework involving
the effect of tangential velocity (caused by magnetic field) and effect
of disorder.

When $B=0$, the flow is radial and the viscous and ohmic dissipation are decoupled in the electric resistance. By solving the Stokes equation, we find
\begin{align}
R\left(0\right) & =R^{\mathsf{vis}}\left(0\right)+R^{\mathsf{ohm}}\left(0\right) \label{eq:R_B=0}\\
% R^{\mathsf{vis}}\left(0\right) & =\frac{\eta_{1}}{\pi\left(n_{1}e\right)^{2}}\left(\frac{1}{r_{1}^{2}}-\frac{1}{r_{\star}^{2}}\right)+\frac{\eta_{2}}{\pi\left(n_{2}e\right)^{2}}\left(\frac{1}{r_{\star}^{2}}-\frac{1}{r_{2}^{2}}\right) \nonumber\\
R^{\mathsf{vis}}\left(0\right) & =\frac{\eta_{1}}{\pi\left(n_{1}e\right)^{2}}\frac{1}{r_{1}^{2}}-\frac{\eta_{2}}{\pi\left(n_{2}e\right)^{2}}\frac{1}{r_{2}^{2}} 
% \nonumber\\ & \hspace{0.5cm} 
+R^{\textsf{vis,int}} \nonumber\\
%
% R^{\textsf{vis,int}} &=\frac{1}{\pi n_{\star}e^{2}r_{\star}^{2}}\left(\frac{\eta_{2}}{n_{2}}-\frac{\eta_{1}}{n_{1}}\right) \nonumber \\
%
% &+ \frac{\eta_{\star}}{\pi n_{\star}e^{2}}\times\frac{\Delta n}{4r_{\star}d} \left(\frac{\eta_{2}}{n_{2}^{2}}-\frac{\eta_{1}}{n_{1}^{2}}\right) \nonumber \\
%
R^{\mathsf{ohm}}\left(0\right) & =\frac{R_{1}^{\mathsf{Dru}}}{2\pi}\ln\frac{r_{\star}}{r_{1}}+\frac{R_{2}^{\mathsf{Dru}}}{2\pi}\ln\frac{r_{2}}{r_{\star}} \nonumber
\end{align}
The first term in $R^{\textsf{vis}}(0)$ is due to the potential drop at the inner lead, the second term is from the outer  lead and $R^{\textsf{vis,int}}$ is due to the potential difference at the interface, determined by Eq. (\ref{eq:BC Phi}). This energy is dissipated by viscous forces in the regions of homogeneous charge density and due to compressive flow at the interface. For parameters in Table \ref{table:parameters} and $n_2=3n_1$, we find the boundary resistance at the inner and outer leads is $0.24\Omega$ while the resistance arising due to potential jump at the interface is $1.76\Omega$, i.e, the total viscous resistance is $R^{\textsf{vis}}(0)=2.0\Omega$. On the other hand, the zero-field ohmic dissipation $R^{\textsf{ohm}}(0)$ is $2.6\Omega$. This conforms well with numerical values of viscous and ohmic dissipation (in Eq. (\ref{eq:power diss})) from our COMSOL simulation ({$2.3\Omega$} and $2.6\Omega$ respectively).

\begin{table*}
\caption{Magneto resistance due to bulk ohmic flow and due to viscous layers, using approximate solution (\ref{eq:vfit}).
Notation: If $\tilde{v}_{i}=-\frac{v\left(r_{i}\right)}{u\left(r_{i}\right)\omega_{c}\tau}$, $i=1,2$ for the inner and outer contacts, respectively, 
and $\tilde{v}_{\star\pm}=-\frac{v_{\star}}{u\left(r_{\star}\pm d\right)\omega_{c}\tau}$, then $\alpha_{1}=\left(\tilde{v}_{1}-1\right)\left(1+\frac{3D_{1}}{2r_{1}}\right)$, $\alpha_{2}=\left(\tilde{v}_{2}-1\right)\left(1-\frac{3D_{2}}{2r_{2}}\right)$,
$\alpha_{\star\mp}=\left(\tilde{v}_{\star\mp}-1\right)\left(1\pm\frac{3D_{1}}{2r_{\star}}\right)$,
$\beta_{i}=\frac{1}{2}\left(\tilde{v}_{i}-1\right)+2\left(\tilde{v}_{i}-1\right)^{2}$, $i=1,2,\star\pm$. $\tilde v_1, \tilde v_2 = 0$ in our simulations, $\tilde v_\star$ is given by Eq. (\ref{vstar}). The expression for $I_1$ is given in Eq. (B1) in appendix.}
\centering
\begin{tabular}{|c|c|c|}
\hline 
$\Delta R\left(B\right)$ & Viscous & Ohmic\tabularnewline
\hline 
\hline 
bulk & $\left(\omega_{c}\tau\right)^{2}\times R^{\mathsf{vis}}\left(0\right)$ & $\left(\omega_{c}\tau\right)^{2}\times R^{\mathsf{ohm}}\left(0\right)$\tabularnewline
\hline 
bdy layer & 
$\begin{array}{c}
\frac{\eta_{1}\left(\omega_{c}\tau\right)^{2}}{\pi\left(n_{1}e\right)^{2}}\left[I_1 +\frac{\alpha_{\star-}^{2}}{4r_{\star}D_1}+\frac{2\alpha_{\star-}}{r_{\star}^{2}}\right]\\
+\frac{\eta_{2}\left(\omega_{c}\tau\right)^{2}}{\pi\left(n_{2}e\right)^{2}}\left[\left(\frac{\alpha_{\star+}^{2}}{4r_{\star}}+\frac{\alpha_{2}^{2}}{4r_{2}}\right)\frac{1}{D_{2}}+\frac{2\alpha_{\star+}}{r_{\star}^{2}}+\frac{2\alpha_{2}}{r_{2}^{2}}\right]
\end{array}$ & $\begin{array}{c}
\frac{\left(\omega_{c}\tau\right)^{2}}{2\pi}\times\bigg[R_{1}^{\mathsf{Dru}}D_{1}\left(\frac{\beta_{1}}{r_{1}}+\frac{\beta_{*-}}{r_{*}}\right)\\
+R_{2}^{\mathsf{Dru}}D_{2}\left(\frac{\beta_{*+}}{r_{*}}+\frac{\beta_{2}}{r_{2}}\right)\bigg]
\end{array}$
\tabularnewline
\hline 
\end{tabular}
\label{table}
\end{table*}

A magnetic field couples the viscous and ohmic dissipation in the electric potential and writing an analytical expression for resistance becomes difficult. However, based on our simplified solution (\ref{eq:vfit}), we can still derive approximate analytic expressions, as summarised in Table \ref{table}. We find that the magnetic field contribution is proportional to $(\omega_c\tau)^2=16$ and is therefore larger than the zero-field resistance ($7.15\Omega$ for viscous and $19.2\Omega$ for ohmic dissipation, from COMSOL). From our simplified expressions, we can quantitatively break up the power dissipation into spatially localized channels. From inner to outer, they are: (i) boundary resistance at the inner lead due to viscous stresses acting on radial flow, (ii) resistance from viscous layer located $\sim D_1$ from the inner lead, arising due to tangential velocity, (iii) potential drop in bulk, primarily due to ohmic scattering, (iv) resistance from viscous layers near the density interface, due to tangential velocity, (v) viscous dissipation due to radial compressive flow at the interface, (vi) bulk resistance due to ohmic scattering in the outer region, followed by the outer viscous layer and the boundary resistance at the outer lead.

\par
\begin{figure}[ht]
    \centering
    \includegraphics[width=0.6\columnwidth]{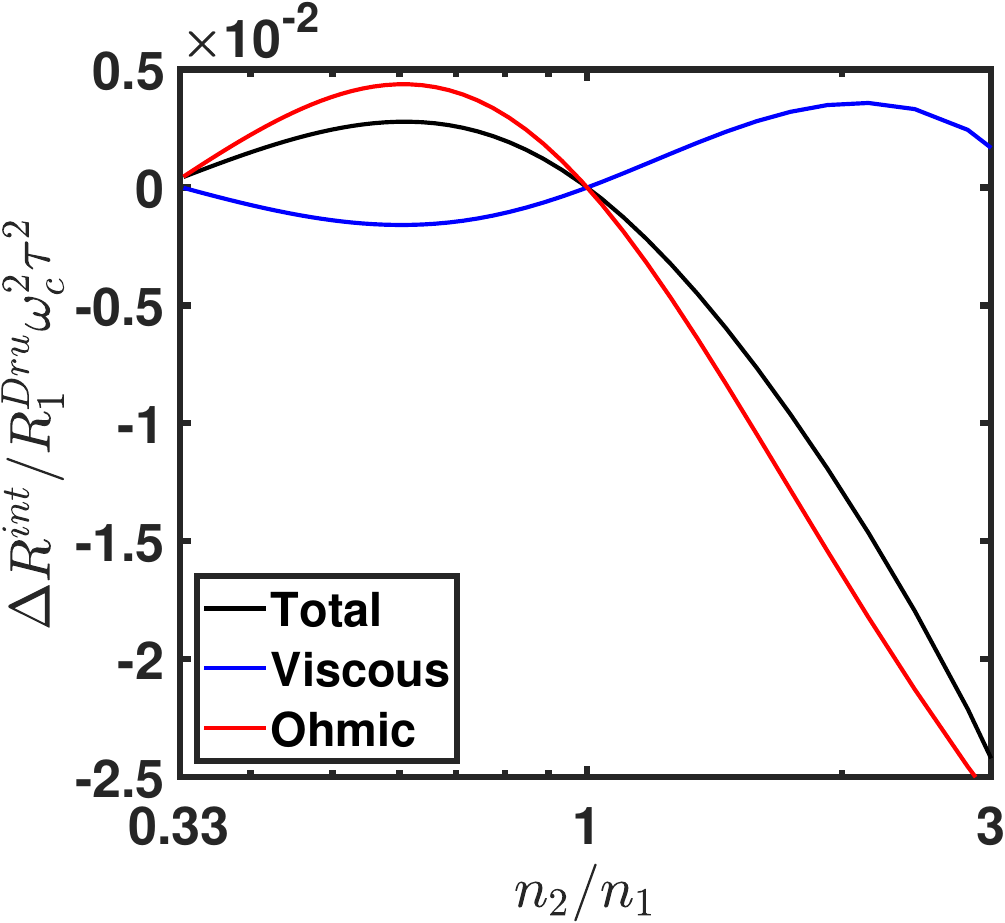}
    \caption{Viscous, ohmic and total contribution to magnetoresistance due to viscous layers at interface, obtained from analytic expressions in table \ref{table}. Negative values indicate that the viscous layers decrease magneto resistance arising from momentum relaxing scattering.
    Because viscosity is proportional to density, viscous dissipation increases with $n_2$ near $n_2/n_1=1$. For high densities, velocity gradients in the viscous layers become small, therefore, viscous dissipation decreases. The opposite is true for very small $n_2/n_1$.
    Also because the viscous layers decelerate the flow more at the interface when $n_2/n_1$ is large, the ohmic dissipation, which is proportional to $n |\mathbf u|^2$, decreases. For $n_2/n_1<1$, the opposite should be true, however, at very low $n_2/n_1$, the decrease in $n$ becomes the overriding factor in ohmic dissipation, and the ohmic contribution starts to fall.
}
    %The total contribution is determined by Ohmic dissipation.}
    \label{fig:Rint}
\end{figure}

 Our expressions in Table \ref{table} allows us to separately calculate the contributions to the total resistance from the different viscous layers, something that would not be possible from either a global measurement or a numerical simulation which probes the total voltage drop and current across the entire device. In Fig (\ref{fig:Rint}), we plot the dissipation from the expressions in Table \ref{table}. We find that for $n_2>n_1$, the net contribution from the interface viscous layers is negative, while for $n_2<n_1$, it is positive. A heuristic explanation can be given as follows. Because viscosity is proportional to density, near $n_2/n_1=1$, viscous dissipation at interface layers increases with $n_2$.  For very high densities, however, the velocity gradients in these layers is small, resulting in decrease in dissipation. The converse happens when $n_2<n_1$. On the other hand, the Ohmic dissipation, which goes as $n|\mathbf u|^2$, drops with increasing density because the velocity of charge carriers decreases. For $n_2/n_1<1$, the increase in $|\mathbf u|$ due to decrease in $n$ saturates, and the Ohmic dissipation starts to decrease due to decrease in $n_2$. The Ohmic contribution dominates over the viscous one, therefore, the net contribution to magneto resistance from the viscous layers is negative when $n_2/n_1>1$ and positive when $n_2/n_1<1$.

\section{Summary}\label{sec:summary}

We have studied hydrodynamic electron flow across density junctions in the Corbino geometry. Starting from the Stokes and continuity equations, we have derived boundary conditions of flow across the interface. Using these conditions, we have shown that we can make tunable viscous layers at the interface by varying the density ratio of the junction. We have also calculated the viscous and Ohmic dissipations by these viscous layers and we have found opposing behaviour when electrons flow from regions of lower or higher density to the other. 

We have based our analysis on experimentally tested values of viscosity and momentum relaxation mean free path at $T/T_F \sim 0.25$ in a 2DEG with parabolic dispersion, as well as a realistic system size.
{It is interesting to note that a novel ``super-Fermi liquid" regime is predicted at $T/T_F \lesssim 0.16$ due to long-lived odd angular modes of the electron distribution function}~\cite{kryhin2023, Ledwith2019}. The theory predicts a linear-in-temperature dependence of conductivity, and an anomalous scaling of local conductivity with characteristic length scale of action of viscous forces~\cite{kryhin2024}. The super-FL theory should be most distinct at low $T$ when the deviation of the decay of long-lived modes of the distribution function from standard Fermi-liquid behaviour is most apparent. At hydrodynamic temperatures considered here ($T/T_F\gtrsim 0.25$), such effects are expected to be mitigated. Hence we conclude that the standard hydrodynamic theory used in this paper suffices to explain our results.

Nevertheless, it would be an interesting idea to explore the effects of the new regime on viscous layers at lower temperatures. The effects of magnetic field and density modulation in the super FL regime is currently an open question and will be the subject of future research.

Density modulation in Corbino using top gates have been achieved in experiments~\cite{Elahi}. Although Ref.~\cite{Elahi} studies the ballistic regime, we can smoothly switch from it to the hydrodynamic regime using temperature as a tuning parameter. This presents the interesting problem of addressing the transport characteristics at this crossover. Study of thermal transport due to viscous layers is another possible avenue of investigation, especially for sample near charge neutrality~\cite{Gall_Corbino_magnetoresistance_in_neutral_graphene,Levchenko_MR}. Additionally, one could thread a magnetic flux through the Corbino like in the quantum Hall setup, as in Ref.~\cite{Tomadin_Corbino_viscometer}. Therefore, our results presented in this paper open up possibilities for exploring novel transport phenomena in the Corbino geometry, and in general in the field of viscous electronic engineering.
\acknowledgments

We are grateful to Giovanni Vignale for valuable discussions and for collaboration on a closely related project. We thank Alexander Hamilton for valuable discussions and experimental insights. This work was supported by the Singapore National Research Foundation Investigator Award (NRF-NRFI06-2020-0003) and the Australian Research Council Centre of Excellence in Future Low-Energy Electronics Technologies (CE170100039).
\appendix
\section{Derivation of approximate solution of Stokes flow}
\label{app:Approximate_solution}

The exact solution of Stokes equation (\ref{eq:Stokes}) is given by (\ref{eq:soln_Stokes}). Here, we derive the simplified form (\ref{eq:approx soln}). Assuming $r_1,r_2\gtrsim D$, we approximate the Bessel functions appearing in the solution by their asymptotic forms. $K_1$ is a decreasing function, hence its contribution towards the outer boundary is small, i.e,
\begin{equation*}
    v\big |_{r\rightarrow r_{2}} \approx c_{1}I_{1}(r/D) +v^{\mathsf{ohm}}(r)\approx c_{1}\frac{e^{r/D}}{\sqrt{2\pi r/D}} +v^{\mathsf{ohm}}(r)
\end{equation*}

Fitting this to the output flow at $r_2$,
\begin{equation*}
   v\big|_{r\rightarrow r_{2}}\approx\left(\omega_c \tau u_2 + v_2\right)\sqrt{\frac{r_{2}}{r}}e^{-(r_{2}-r)/D}  -\omega_c\tau \frac{u_2r_2}{r}
\end{equation*}

We find that viscous correction to the Ohmic flow predominates in a region of width
$\approx D$ from the outer boundary.

\par Similarly, because $I_{1}$ is an increasing function of $r$, and given that inner and outer boundaries are largely separated, we expect the relative contribution of $I_{1}$ towards inner boundary is small.
\begin{equation*}
    v\big|_{r\rightarrow r_{1}}\approx c_2K_1(r/D) +v^{\mathsf{ohm}}(r) \approx c_{2}\sqrt{\frac{\pi D}{2r}}e^{-r/D} +v^{\mathsf{ohm}}(r)
\end{equation*}

Fitting to input flow at $r_{1}$:
\begin{equation*}
    v\big|_{r\rightarrow r_{1}}\approx\left(\omega_c\tau u_1 + v_1\right)\sqrt{\frac{r_{1}}{r}}e^{-(r-r_{1})/D} -\omega_c\tau\frac{u_1r_1}{r}
\end{equation*}

Stitching these together in (\ref{eq:soln_Stokes}), we get the required expression.

\par Although the above approximations hold for $D\lesssim r_1$, a similar approach can be made for $r_1 \lesssim D$ ($r_2$ large). This situation, seemingly impractical, is now a possibility with the fabrication of ultra-clean semiconductor heterojunctions. In recent experiments, mobilities as high as $50\times 10^6 \text{cm}^2/\text V\cdot \text s$ have been reached~\cite{Pfieffer_ultra_high_mob_2DES}, for which the Gurzhi length is of the order of $10\mu m$. To simplify the Stokes solution, we assume a hypothetical interface at $r_{\star}=r_1+D$ and divide the solution into two regions. The outer region, similar to before, has the form

\begin{equation}\label{eq:v_out}
\begin{split}
    v^{\mathsf{out}}\approx&\left(\omega_c\tau u_{\star} +v_{\star}\right)\sqrt{\frac{r_{\star}}{r}}e^{-(r-r_{\star})/D} -\omega_c\tau\frac{u_1r_{1}}{r}\\
    &+\left(\omega_c\tau u_2 + v_2\right)\sqrt{\frac{r_{2}}{r}}e^{-(r_{2}-r)/D}
\end{split}
\end{equation}

For the inner region $r_1<r<r_{\star}$, we make small-argument expansion of the Bessel functions in the exact solution (\ref{eq:soln_Stokes}):
\[
v^{\mathsf{in}}=A\frac{r}{r_{1}}+A'\frac{r_{1}}{r}+A'\frac{r_{1}}{2}r\ln\frac{r}{r_{1}} + O(r^2/D^2)
\]
where coefficients $A,A^\prime$ have been suitably defined to impose boundary conditions. Matching the flow at the input terminal,

\begin{equation}\label{eq:v_in}
\begin{split}
    v^{\mathsf{in}}&=\left(v_{1}-A\right)\frac{r_{1}}{r}+A\frac{r}{r_{1}}
    +\frac{r_{1}}{2D^2} \left(v_{1}-A+\omega_c\tau u_1 \right) r\ln\frac{r}{r_{1}}
\end{split}
\end{equation}

\begin{figure}[h]
    \centering
    \includegraphics[width=0.6\columnwidth]{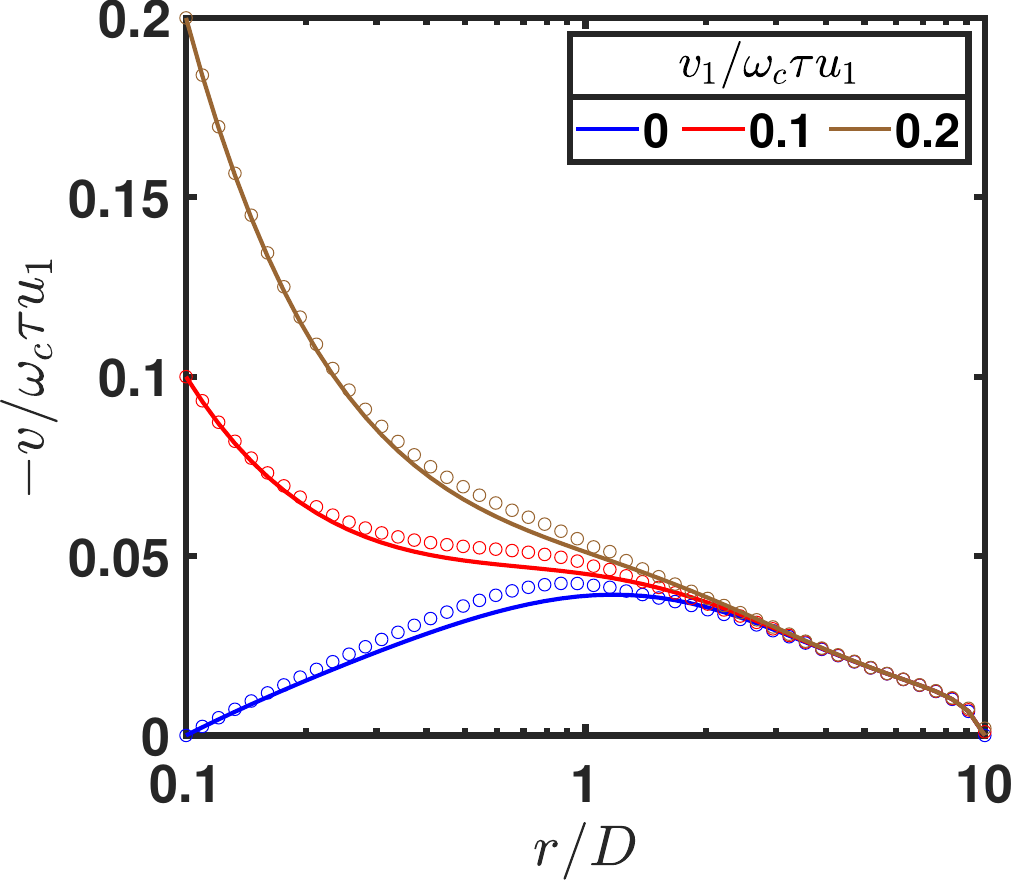}
    \caption{Approximate solution of Stokes flow (circles) for different injected velocities when $r_1\lesssim D$. Solid lines represent the exact solution}
    \label{fig:vfit_r1_lesssim_D}
\end{figure}

The remaining unknown $A$ can be determined numerically by imposing continuity of vorticity at $r_{\star}$. % This gives
% \begin{equation*}
%     \begin{split}
%         A&=-(v_1 + \omega_c\tau u_1)\times\\
%         &\frac{\left(r_{*}+\frac{1}{2}\right)\sqrt{\frac{r_{2}}{r}}e^{-r_{2}+r_{*}}+\frac{1}{2}-r_{*}-\frac{r_{*}^{2}}{2}+\left(r_{*}+\frac{3}{2}\right)2r_{*}^{2}\ln\frac{r_{*}}{r_{1}}}{\left(r_{*}+\frac{3}{2}\right)\left(2r_{*}^{2}\ln\frac{r_{*}}{r_{1}}-\frac{4r_{*}^{2}}{r_{1}^{2}}\right)-\left(\frac{1}{2}-r_{*}-\frac{r_{*}^{2}}{2}\right)}
%     \end{split}
% \end{equation*}
%
The complete solution (\ref{eq:v_out}), (\ref{eq:v_in}) is plotted in Fig (\ref{fig:vfit_r1_lesssim_D}) for the hypothetical case of different injected tangential velocities. The approximate analytic solution is also compared with the exact solution (\ref{eq:soln_Stokes}).

%%%%%%

\section{Derivation of magnetoresistance expressions in Table \ref{table}}

\subsection{Ohmic dissipation}

From the power budget equation (\ref{eq:power diss}), the Ohmic power dissipation
is
\[
P^{\mathsf{ohm}}=2\pi m^{*}\tau^{-1}\int rn\left(u^{2}+v^{2}\right)dr
\]

First, let us consider homogeneous charge density. The radial component
is simply $u=u_{1}r_{1}/r$, hence, the corresponding contribution
to $P^{\mathsf{ohm}}$ is
\[
P^{\mathsf{ohm},r}=I^{2}\frac{R^{\mathsf{Dru}}}{2\pi}\ln\frac{r_{2}}{r_{1}}
\]
\\
where the total current $I=2\pi r_{1}neu_{1}$ and $R^{\mathsf{Dru}}=m^{*}/ne^{2}\tau$.

From approximate solution (\ref{eq:approx soln}), we can split the tangential velocity
into a bulk ohmic contribution and a boundary contribution as
\[
v=v^{\mathsf{ohm}}+v^{\mathsf{bdy}}
\]
\\
where $v^{\mathsf{bdy}}\propto e^{-\left(r-r_{1}\right)/D}$ near $r_{1}$
and $v^{\mathsf{bdy}}\propto e^{-\left(r_{2}-r\right)/D}$ near $r_{2}$. From
this, the tangential flow contribution to $P^{\mathsf{ohm}}$ is
\begin{align*}
P^{\mathsf{ohm},t}=&\frac{2\pi m^{*}n}{\tau}\bigg\{ \int r\left(v^{\mathsf{ohm}}\right)^{2}dr\\
&+\int r\left[2v^{\mathsf{ohm}}v^{\mathsf{bdy}}+\left(v^{\mathsf{bdy}}\right)^{2}\right]dr\bigg\} 
\end{align*}
\\
The first term, which is an ohmic contribution and comes from the
bulk, is:
\begin{align*}
P^{\mathsf{ohm},t}\bigg|_{\mathsf{bulk}}&=I^{2}\left(\omega_{c}\tau\right)^{2}\frac{R^{\mathsf{Dru}}}{2\pi}\ln\frac{r_{2}}{r_{1}}\\
&=\left(\omega_{c}\tau\right)^{2}P^{\mathsf{ohm},r}=\left(\omega_{c}\tau\right)^{2}P^{\mathsf{ohm}}\bigg|_{B=0}
\end{align*}
\\
The remaining contribution, coming from the viscous boundary layers,
is
\begin{align*}
P^{\mathsf{ohm},t} &\bigg|_{\mathsf{inner\;bdy}}\times\frac{\tau}{2\pi m^{*}n} \\
& =-2\omega_{c}\tau u_{1}r_{1}\left(v_{1}+\omega_{c}\tau u_{1}\right)\int_{r_{1}}^{\infty}e^{-\left(r-r_{1}\right)/D}dr\\
 & +r_{1}\left(v_{1}+\omega_{c}\tau u_{1}\right)^{2}\int_{r_{1}}^{\infty}e^{-2\left(r-r_{1}\right)/D}dr\\
 & =\left(\omega_{c}\tau u_{1}\right)^{2}r_{1}D\left\{ 2\left(\tilde{v}_{1}-1\right)+\frac{1}{2}\left(\tilde{v}_{1}-1\right)^{2}\right\} 
\end{align*}
with the notation $\tilde{v}_{1}=-v_{1}/\omega_{c}\tau u_{1}$. An
exactly similar calculation gives
\begin{align*}
P^{\mathsf{ohm},t} &\bigg|_{\mathsf{outer\;bdy}}\times\frac{\tau}{2\pi m^{*}n} \\
&=\left(\omega_{c}\tau u_{2}\right)^{2}r_{2}D\left\{ 2\left(\tilde{v}_{2}-1\right)+\frac{1}{2}\left(\tilde{v}_{2}-1\right)^{2}\right\} 
\end{align*}
with $u_{2}=u\left(r_{2}\right)=u_{1}r_{1}/r_{2}$, and $\tilde{v}_{2}=-v_{2}/\omega_{c}\tau u_{2}$.

The generalization to a system with two different densities is straightforward:
the power dissipated is the sum of the dissipations from each of the
uniform density regions. The result is given in Table \ref{table}.

\subsection{Viscous dissipation}
% \label{appendix:visc_diss}
\[
P^{\mathsf{vis}}=2\pi\int\frac{\boldsymbol{\sigma}^{2}}{2\eta}r\;dr
\]
Under azimuthal symmetry, $\sigma_{rr}=-\sigma_{\phi\phi}=\eta r\partial_{r}\left(u/r\right)$, while $\sigma_{r\phi}=\sigma_{\phi r}=\eta r\partial_{r}\left(v/r\right)$.
The viscous dissipation due to radial flow
\[
P^{\mathsf{vis},r}=\frac{2\pi}{\eta}\int\sigma_{rr}^{2}r\;dr=I^{2}\frac{\eta}{\pi\left(ne\right)^{2}}\left(\frac{1}{r_{1}^{2}}-\frac{1}{r_{2}^{2}}\right)
\]
This is the same as power supplied due to boundary stresses in the
clean limit, when the electric field driving the current disappears~\cite{Shavit}. The magnetic field correction, and also the contribution
from disorder, comes from the azimuthal flow, as
\begin{align*}
P^{\mathsf{vis},t} & =\frac{2\pi}{\eta}\int\sigma_{r\phi}^{2}r\;dr\\
 & =2\pi\eta\bigg\{\int_{r_{1}}^{r_{2}}r^{3}\left[\partial_{r}\left(\frac{v^{\textsf{ohm}}}{r}\right)\right]^{2}dr\\
 & +\int_{r_{1}}^{r_{2}}r^{3}\left[2\partial_{r}\left(\frac{v^{\textsf{ohm}}}{r}\right)\partial_{r}\left(\frac{v^{\mathsf{bdy}}}{r}\right)+\partial_{r}\left(\frac{v^{\textsf{bdy}}}{r}\right)^{2}\right]dr\bigg\}
\end{align*}
Like for ohmic dissipation, the first term is proportional to the
zero $B$ field (bulk) resistance
\[
P^{\mathsf{vis},t}\bigg|_{\mathsf{bulk}}=\left(\omega_{c}\tau\right)^{2}P^{\mathsf{vis},r}
\]
The contribution from the boundary layers comes from the second term.
At the inner boundary layer,
\begin{align*}
P^{\mathsf{vis},t}\bigg|_{\mathsf{inner\;bdy}} & =I^{2}\frac{\eta (\omega_c\tau)^2}{\pi\left(ne\right)^{2}}\bigg\{\int_{r_{1}}^{\infty}F_{1}\left(r\right)e^{-\left(r-r_{1}\right)/D}dr\\
 & +\int_{r_{1}}^{\infty}F_{2}\left(r\right)e^{-2\left(r-r_{1}\right)/D}dr\bigg\}\\
F_{1}\left(r\right) & =\frac{4}{r_{1}^{1/2}r^{3/2}}\left(\tilde{v}_{1}-1\right)\left(1+\frac{3D}{2r}\right)\\
F_{2}\left(r\right) & =\frac{1}{r_{1}}\left(\tilde{v}_{1}-1\right)^{2}\left(1+\frac{3D}{2r}\right)^{2}
\end{align*}
with $\tilde{v}_{1}=-v_{1}/\omega_{c}\tau u_{1}$ as before. If $r_{1}\gg D$,
$F_{1},F_{2}$ are slowly varying with $r$ compared to the exponential,
hence, we may set their values as fixed at $r_{1}$.
\begin{align*}
P^{\mathsf{vis},t}\bigg|_{\mathsf{inner\;bdy}} & =I^{2}\frac{\eta}{\pi\left(ne\right)^{2}}\left\{ \frac{2\alpha_{1}}{r_{1}^{2}}+\frac{\alpha_{1}^{2}}{4r_{1}D}\right\} \\
\alpha_{1} & =\left(\tilde{v}_{1}-1\right)\left(1+\frac{3D}{2r_{1}}\right)
\end{align*}

A very similar expression holds at $r_{2}$:
\begin{align*}
P^{\mathsf{vis},t}\bigg|_{\mathsf{outer\;bdy}} & =I^{2}\frac{\eta (\omega_c\tau)^2}{\pi\left(ne\right)^{2}}\left\{ \frac{2\alpha_{2}}{r_{2}^{2}}+\frac{\alpha_{2}^{2}}{4r_{2}D}\right\} \\
\alpha_{2} & =\left(\tilde{v}_{2}-1\right)\left(-1+\frac{3D}{2r_{1}}\right)
\end{align*}

The dissipation due to two density regions is the sum of dissipation
from each region; however, for our choice of parameters $r_{1}=4D$,
the assumption $r_{1}/D\gg1$ is not valid and naively using the above
expression gives inaccurate estimates near $r_{1}$. In this case,
we directly integrate the full expression keeping the generic form
of $F_{1}\left(r\right),F_{2}\left(r\right)$. The result can be expressed
in terms of the exponential integral function:
% \begin{align*}
% P^{\mathsf{vis},t}\bigg|_{\mathsf{inner\;bdy}}	=&I^{2}\frac{\eta}{\pi\left(ne\right)^{2}}\bigg\{ \frac{2\left(\tilde{v}_{1}-1\right)}{r_{1}^{2}} \\
% &+\frac{\left(\tilde{v}_{1}-1\right)^{2}}{4r_{1}D}\left(1+\frac{3D}{2r_{1}}\right)+\lambda\bigg\} \\
% \lambda	=&\frac{3}{4r_{1}^{2}}\left(\tilde{v}_{1}-1\right)^{2}\left(1+\frac{r_{1}}{D}e^{2r_{1}/D}\mathrm{Ei}\left[-\frac{2r_{1}}{D}\right]\right) \\
% \mathrm{Ei}\left[z\right]	=&-\int_{-z}^{\infty}dt\;e^{-t}/t
% \end{align*}

\begin{gather} \label{eq:appendix-I2}
    P^{\mathsf{vis},t}\bigg|_{\mathsf{inner\;bdy}}=  I^{2}\frac{\eta\left(\omega_{c}\tau\right)^{2}}{\pi\left(ne\right)^{2}}I_{1} \nonumber \\
I_{1}= \bigg\{\frac{2\left(\tilde{v}_{1}-1\right)}{r_{1}^{2}}+\frac{\left(\tilde{v}_{1}-1\right)^{2}}{4r_{1}D}\left(1+\frac{3D}{2r_{1}}\right)+\lambda\bigg\}\\
\lambda=  \frac{3}{4r_{1}^{2}}\left(\tilde{v}_{1}-1\right)^{2}\left(1+\frac{r_{1}}{D}e^{2r_{1}/D}\mathrm{Ei}\left[-\frac{2r_{1}}{D}\right]\right) \nonumber\\
\mathrm{Ei}\left[z\right]=  -\int_{-z}^{\infty}dt\;e^{-t}/t \nonumber
\end{gather}

\bibliographystyle{apsrev4-2}
\bibliography{revised_manuscript.bib}

\end{document}